\documentclass[prl,a4paper,twocolumn]{revtex4}%
\usepackage{amsfonts}
\usepackage{amsmath}
\usepackage{amssymb}
\usepackage{graphicx}%
\begin{document}
\title[Octonionic gravitation]{Natural octonionic generalization of general relativity}
\author{John Fredsted}
\email{physics@johnfredsted.dk}
\affiliation{Soeskraenten 22, Stavtrup, DK-8260 Viby J., Denmark}


\begin{abstract}
An intriguingly natural generalization, using complex octonions, of general
relativity is pointed out. The starting point is the vierbein-based double
dual formulation of the Einstein-Hilbert action. In terms of two natural
structures on the (complex) quaternions and (complex) octonions, the inner
product and the cross products, respectively, this action is linked with the
complex quaternionic structure constants, and subsequently generalized to an
achtbein-based 'double $\chi$-dual' action in terms of the complex octonionic
structure constants.

\end{abstract}
\maketitle

It is the purpose of this note to point out an intriguingly natural
generalization of general relativity, using complex octonionic structures.

In terms of the doubly contracted, double dual \cite[p. 325]{MTW} of the
Riemann curvature tensor $R_{\mu\nu}{}^{\rho\sigma}$, the Einstein-Hilbert
action may be written as%
\begin{align*}
S_{\mathrm{EH}}  &  =\int L_{\mathrm{EH}}\sqrt{-g}d^{4}x,\\
L_{\mathrm{EH}}  &  =-\frac{c^{4}}{64\pi G}\varepsilon^{\alpha\beta\mu\nu
}R_{\mu\nu}{}^{\rho\sigma}\varepsilon_{\rho\sigma\alpha\beta},
\end{align*}
where $g\equiv\det\left(  g_{\mu\nu}\right)  $ is the determinant of the
metric, and $\varepsilon_{\mu\nu\rho\sigma}$ and $\varepsilon^{\mu\nu
\rho\sigma}$ are the Levi-Civita tensor densities \cite[Eq. (8.10)]{MTW}%
\begin{align*}
\varepsilon_{\mu\nu\rho\sigma}  &  =+\left(  -g\right)  ^{+1/2}\left[  \mu
\nu\rho\sigma\right]  ,\\
\varepsilon^{\mu\nu\rho\sigma}  &  =-\left(  -g\right)  ^{-1/2}\left[  \mu
\nu\rho\sigma\right]  ,
\end{align*}
where $\left[  \mu\nu\rho\sigma\right]  $ is a completely antisymmetric symbol
with $\left[  0123\right]  =+1$. This assertion is an immediate consequence of
the identity%
\[
-\frac{1}{2}\varepsilon^{\alpha\beta\mu\nu}\varepsilon_{\alpha\beta\rho\sigma
}=\delta_{\rho}^{\mu}\delta_{\sigma}^{\nu}-\delta_{\sigma}^{\mu}\delta_{\rho
}^{\nu}\equiv\delta_{\rho\sigma}^{\mu\nu},
\]
where $\delta_{\rho\sigma}^{\mu\nu}$ is a generalized Kronecker delta
\cite[Sect. 4.2]{Lovelock and Rund}.

In terms of a vierbein $e^{a}{}_{\mu}$, with corresponding minimal spin
connection $\omega_{\mu}{}^{ab}\equiv g^{\rho\sigma}e^{a}{}_{\rho}\nabla_{\mu
}e^{b}{}_{\sigma}$, the Einstein-Hilbert action may equivalently be written
as
\begin{align*}
S &  =\int Led^{4}x,\\
L &  =-\frac{c^{4}}{64\pi G}\varepsilon^{\alpha\beta\mu\nu}e^{\rho}{}%
_{a}e^{\sigma}{}_{b}R_{\mu\nu}{}^{ab}\varepsilon_{\rho\sigma\alpha\beta},
\end{align*}
where $e\equiv\det\left(  e^{a}{}_{\mu}\right)  $ is the determinant of the
vierbein, $R_{\mu\nu}{}^{ab}$ is the curvature tensor of the minimal spin
connection \cite{GSW,Weinberg}, and $\varepsilon_{\mu\nu\rho\sigma}$ and
$\varepsilon^{\mu\nu\rho\sigma}$ are now given by%
\begin{align*}
\varepsilon_{\mu\nu\rho\sigma} &  =e^{a}{}_{\mu}e^{b}{}_{\nu}e^{c}{}_{\rho
}e^{d}{}_{\sigma}\varepsilon_{abcd},\\
\varepsilon^{\mu\nu\rho\sigma} &  =e^{\mu}{}_{a}e^{\nu}{}_{b}e^{\rho}{}%
_{c}e^{\sigma}{}_{d}\varepsilon^{abcd},
\end{align*}
because $\varepsilon_{abcd}=-\varepsilon^{abcd}=\left[  abcd\right]  $ in any
Minkowski frame.

By itself, all of the above would of course be quite pointless, were it not
for the following fact: The above vierbein-based double dual action allows for
an intriguingly natural (i.e., uncontrived) generalization, in terms of
complex octonions, of general relativity, using the complex quaternions as a
stepping-stone. In order to analytically formulate this assertion, two
structures, the inner product and the triple cross products, see shortly, must
first be defined.

Remark:\ In this note, no formal introduction to the (complex) quaternions or
(complex) octonions will be given. Instead, the reader is kindly referred to
the literature: For short reviews of the octonions, see for instance Refs.
\cite{Gunaydin and Gursey,Dundarer and Gursey,Dundarer Gursey and Tze,Bakas et
al.}. For a comprehensive review of the octonions, see Ref. \cite{Baez}. For a
monograph on octonions and other nonassociative algebras, see Ref.
\cite{Okubo}. In particularly, for a monograph on composition algebras, a
class to which both the complex quaternions and complex octonions belong (note
that they are not division algebras, even though the quaternions and the
octonions themselves are), see Ref. \cite{Springer and Veldkamp}. For material
on the quaternions, see for instance Refs. \cite{Baez,Okubo,Lambek}.

Below, $\mathbb{D}$ (for division algebra) denotes either the set of
quaternions $\mathbb{H}$, or the set of octonions $\mathbb{O}$. Standardly,
although it differs in various references by a normalizing factor of $2$,
define an inner product $\left\langle \cdot,\cdot\right\rangle :\left(
\mathbb{C}\otimes\mathbb{D}\right)  ^{2}\rightarrow\mathbb{C}$ by%
\[
2\left\langle x,y\right\rangle =x\overline{y}+y\overline{x}\equiv\overline
{x}y+\overline{y}x,
\]
where $\overline{x}$ denotes the (quaternionic or octonionic) conjugate of
$x$. Define triple cross products $X_{L},X_{R}:\left(  \mathbb{C}%
\otimes\mathbb{D}\right)  ^{3}\rightarrow\mathbb{C}\otimes\mathbb{D}$ by%
\begin{align*}
3!X_{L}\left(  x,y,z\right)   &  =x\left(  \overline{y}z-\overline{z}y\right)
+\text{cyclic perm},\\
3!X_{R}\left(  x,y,z\right)   &  =\left(  x\overline{y}-y\overline{x}\right)
z+\text{cyclic perm},
\end{align*}
where $L$ and $R$ means left and right, respectively. The cross products
$X_{L}$ and $X_{R}$ possess both the orthogonality property and the
(generalized) Pythagorean property \cite{Lounesto};%
\begin{align*}
0  &  =\left\langle X\left(  x_{1},x_{2},x_{3}\right)  ,x_{i}\right\rangle ,\\
\det\left(  \left\langle x_{i},x_{j}\right\rangle \right)   &  =\left\langle
X\left(  x_{1},x_{2},x_{3}\right)  ,X\left(  x_{1},x_{2},x_{3}\right)
\right\rangle ,
\end{align*}
where the suppressed subscript means that the relations apply to both $L$ and
$R$. Trilinear cross products possessing both these properties exist only over
algebras of real (or complex) dimension $4$ or $8$ \cite{Lounesto,Zvengrowski}%
, the underlying reason being the existence of precisely the division algebras
$\mathbb{H}$ and $\mathbb{O}$.

Remark: For $\mathbb{D}=\mathbb{H}$, the cross products $X_{L}$ and $X_{R}$
are trivially identical because the quaternions are associative, so for the
quaternions the subscript will be dropped, writing just $X$ (for both). For
$\mathbb{D}=\mathbb{O}$, however, this is not the case because the octonions
are nonassociative.

As bases (over $\mathbb{C}$) for $\mathbb{C}\otimes\mathbb{H}$ and
$\mathbb{C}\otimes\mathbb{O}$, respectively, define $e_{a}=\left(
\mathrm{i},e_{i}\right)  $ and $E_{a}=\left(  \mathrm{i},E_{i}\right)  $,
where $\mathrm{i}$ is the complex imaginary unit, $e_{i}$ are the standard
units of the pure quaternions $\operatorname{Im}\mathbb{H}$ (quaternions with
no real part), and $E_{i}$ are the standard units of the pure octonions
$\operatorname{Im}\mathbb{O}$ (octonions with no real part). The units $e_{i}$
obey%
\begin{align*}
2\varepsilon_{ij}{}^{k}e_{k}  &  =\left[  e_{i},e_{j}\right]  ,\\
0  &  =\left[  e_{i},e_{j},e_{k}\right]  ,
\end{align*}
where $\varepsilon_{ijk}$ are the completely antisymmetric structure constants
of the quaternions (with $\varepsilon_{123}=+1$). The units $E_{i}$ obey%
\begin{align*}
2\psi_{ij}{}^{k}E_{k}  &  =\left[  E_{i},E_{j}\right]  ,\\
2\phi_{ijk}{}^{l}E_{l}  &  =\left[  E_{i},E_{j},E_{k}\right]  ,
\end{align*}
where $\psi_{ijk}\in\mathbb{R}$ and $\phi_{ijkl}\in\mathbb{R}$ (each others
dual in $\mathbb{R}^{7}$) are the completely antisymmetric structure constants
of the octonions. Above, $\left[  \cdot,\cdot\right]  :\left(  \mathbb{C}%
\otimes\mathbb{D}\right)  ^{2}\rightarrow\mathbb{C}\otimes\mathbb{D}$ and
$\left[  \cdot,\cdot,\cdot\right]  :\left(  \mathbb{C}\otimes\mathbb{D}%
\right)  ^{3}\rightarrow\mathbb{C}\otimes\mathbb{D}$, defined by%
\begin{align*}
\left[  x,y\right]   &  =xy-yx,\\
\left[  x,y,z\right]   &  =\left(  xy\right)  z-x\left(  yz\right)  ,
\end{align*}
are the commutator and associator, respectively.

Remark: The following index conventions are adhered to throughout. Lower case
Latin letters from the middle of the alphabet, starting at $i$, either run
from $1$ to $3$, or from $1$ to $7$, depending on the context. They are raised
and lowered with $\delta^{ij}$ and $\delta_{ij}$, respectively. Lower case
Latin letters from the beginning of the alphabet either run from $0$ to $3$,
or from $0$ to $7$, depending on the context. They are raised and lowered with
$\eta^{ab}$ and $\eta_{ab}$, respectively. Lower case greek letters either run
from $0$ to $3$, or from $0$ to $7$, depending on the context. They are raised
and lowered with $g^{\mu\nu}$ and $g_{\mu\nu}$, respectively.

Now, the seemingly insignificant but in fact crucial observation is that%
\[
\varepsilon_{abcd}=\mathrm{i}\left\langle X\left(  e_{a},e_{b},e_{c}\right)
,e_{d}\right\rangle ,
\]
a relation linking duality in spacetime, as controlled by $\varepsilon_{abcd}%
$, with two natural structures of the (complex) quaternions, the inner product
and the cross product. This relation is the complex quaternionic
stepping-stone, previously referred to.

Because the (complex) quaternions can be embedded in the (complex) octonions
(in numerous ways), it seems natural to generalize as follows: Define
$\chi_{abcd}^{\left(  L\right)  }$ and $\chi_{abcd}^{\left(  R\right)  }$
(eight-dimensional generalizations of the four-dimensional $\varepsilon
_{abcd}$) by%
\begin{align*}
\chi_{abcd}^{\left(  L\right)  }  &  =\mathrm{i}\left\langle X_{L}\left(
E_{a},E_{b},E_{c}\right)  ,E_{d}\right\rangle ,\\
\chi_{abcd}^{\left(  R\right)  }  &  =\mathrm{i}\left\langle X_{R}\left(
E_{a},E_{b},E_{c}\right)  ,E_{d}\right\rangle .
\end{align*}
Even though not obvious, $\chi_{abcd}^{\left(  L\right)  }$ and $\chi
_{abcd}^{\left(  R\right)  }$ are in fact completely antisymmetric.
Subsequently, define $\chi_{\mu\nu\rho\sigma}^{\left(  L\right)  }$ and
$\chi_{\mu\nu\rho\sigma}^{\left(  R\right)  }$ (eight-dimensional
generalizations of the four-dimensional $\varepsilon_{\mu\nu\rho\sigma}$) by%
\begin{align*}
\chi_{\mu\nu\rho\sigma}^{\left(  L\right)  }  &  =E^{a}{}_{\mu}E^{b}{}_{\nu
}E^{c}{}_{\rho}E^{d}{}_{\sigma}\chi_{abcd}^{\left(  L\right)  },\\
\chi_{\mu\nu\rho\sigma}^{\left(  R\right)  }  &  =E^{a}{}_{\mu}E^{b}{}_{\nu
}E^{c}{}_{\rho}E^{d}{}_{\sigma}\chi_{abcd}^{\left(  R\right)  },
\end{align*}
where $E^{a}{}_{\mu}$ is an achtbein. Finally, define an achtbein-based
'double $\chi$-dual' action (generalization of the vierbein-based double dual
action) by
\begin{align*}
S  &  =\int LEd^{8}x,\\
L  &  =-\frac{c^{4}}{64\pi G}\chi_{\left(  L\right)  }^{\alpha\beta\mu\nu
}E^{\rho}{}_{a}E^{\sigma}{}_{b}R_{\mu\nu}{}^{ab}\chi_{\rho\sigma\alpha\beta
}^{\left(  R\right)  },
\end{align*}
where $E\equiv\det\left(  E^{a}{}_{\mu}\right)  $ is the determinant of the
achtbein, and $R_{\mu\nu}{}^{ab}$ is the curvature tensor of the associated
minimal spin connection $\Omega_{\mu}{}^{ab}\equiv g^{\rho\sigma}E^{a}{}%
_{\rho}\nabla_{\mu}E^{b}{}_{\sigma}$. This is the generalized action promised
in this note.

By straightforward calculations, the following relations may be obtained, the
remaining components following from the complete antisymmetry of both
$\chi_{abcd}^{\left(  L\right)  }$ and $\chi_{abcd}^{\left(  R\right)  }$:%
\begin{align*}
\chi_{0ijk}^{\left(  L\right)  }  &  =+\chi_{0ijk}^{\left(  R\right)  }%
=\psi_{ijk},\\
\chi_{ijkl}^{\left(  L\right)  }  &  =-\chi_{ijkl}^{\left(  R\right)
}=\mathrm{i}\phi_{ijkl}.
\end{align*}
These relations show that $\chi_{abcd}^{\left(  L\right)  }$ and $\chi
_{abcd}^{\left(  R\right)  }$, and therefore also $\chi_{\mu\nu\rho\sigma
}^{\left(  L\right)  }$ and $\chi_{\mu\nu\rho\sigma}^{\left(  R\right)  }$,
are each others complex conjugate, implying that the above Lagrangian is
real-valued, as befits any reasonable Lagrangian.

\end{document}